# Theoretical and Numerical Analyses of a Slit-Masked Chicane for Modulated Bunch Generation

**Xiaofang Zhu**[a*], **Daniel R Broemmelsiek**[b], **and Young-Min Shin**[a, b†]

[a] *Department of Physics, Northern Illinois University, Dekalb, IL, 60115, USA*
[b] *Fermi National Accelerator Laboratory, Batavia, IL, USA*
  *E-mail*: yshin@niu.edu

ABSTRACT: Density modulations on electron beams can improve machine performance of beam-driven accelerators and FELs with resonance beam-wave coupling. The beam modulation is studied with a masked chicane by the analytic model and simulations with the beam parameters of the Fermilab Accelerator Science and Technology (FAST) facility. With the chicane design parameters (bending angle of 18°, bending radius of 0.95 m and $R_{56}$ ~ - 0.19 m) and a nominal beam of 3 ps bunch length, the analytic model showed that a slit-mask with slit period 900 μm and aperture width 300 μm induces a modulation of bunch-to-bunch spacing ~100 μm to the bunch with 2.4% correlated energy spread. With the designed slit mask and a 3 ps bunch, particle-in-cell (PIC) simulations, including nonlinear energy distributions, space charge force, and coherent synchrotron radiation (CSR) effect, also result in beam modulation with bunch-to-bunch distance around 100 μm and a corresponding modulation frequency of 3 THz. The beam modulation has been extensively examined with three different beam conditions, 2.25 ps (0.25 nC), 3.25 ps (1 nC), and 4.75 ps (3.2 nC)，by tracking code Elegant. The simulation analysis indicates that the sliced beam by the slit-mask with 3 ~ 6% correlated energy spread has modulation lengths about 187 μm (0.25 nC), 270 μm (1 nC) and 325 μm (3.2 nC). The theoretical and numerical data proved the capability of the designed masked chicane in producing modulated bunch train with micro-bunch length around 100 fs.

---

[*] Current address: School of Physical Electronics, UESTC, Chengdu, Sichuan, 610054, P. R. China
[†] Corresponding author.

**Contents**

## 1. Introduction

A wide range of electron beam applications such as free-electron lasers (FELs) and particle accelerators employ a bunched beam for improvement of machine performance [1 – 9] in the relativistic or quasi-relativistic regimes. It is well known that a short electron pulse can lead to an appreciable improvement of energy conversion efficiencies or power growth of coherent light sources and high gradient accelerators. Generally, electron bunches can be shortened by a magnetic bunch compressor (e.g. chicane, S-chicane, dogleg, $\alpha$-magnet) [1, 7, 16 – 18] or continuous velocity bunching process [10, 11]. Although the bunching techniques are commonly used in electron accelerators, it is not easy to reduce the bunch length to sub-ps range. Normally, when a beam is modulated, it is more strongly coupled with an undulating or accelerating structure at the resonance condition with a fundamental or higher order mode [19, 20]. The bunch modulation would enable more sophisticated beam control in energy-phase space. The modulation (or microbunching) thus enables photo-electron interaction or electro-optical transition in a smaller time scale, preferably in femto-second range [12]. One of the easiest ways to achieve the beam-modulation is to mask the beam in a chicane with a slit-mask or a wire-grid. The basic concept was first suggested by D. C. Nguyen and B. Carlsten in 1996 in the effort to reduce the length of FEL undulators [4, 5]. Also, the Brookhaven National Laboratory (BNL) demonstrated the generation of a stable train of micro-bunches with a controllable sub-ps delay with the mask technique using a wire-grid [6]. The main advantage of the masking technique is to readily control micro-structured density profiles, including the energies and phases.

We have been investigating the masked chicane technique with the available beam parameters such as the 50 MeV photoinjector of the Fermilab Accelerator Science and Technology (FAST) facility, which is currently being constructed and commissioned in Fermilab [7]. Downstream of the FAST 50 MeV photoinjector beamline, a magnetic bunch compressor, consisting of four rectangular dipoles, is adopted and a slit-mask is designed and inserted in the middle (Fig. 1). Based on this slit-masked chicane, the bunching performance and the ability of sub-ps microbunch generation are studied. In order to evaluate bunching performance with nominal beam parameters, the masked chicane has been analyzed by the linear bunching theory in terms of bunch-to-bunch distance and microbunch length. Two simulation codes, CST-PS [13] and Elegant [14, 15], are employed to examine the theoretical model with the FAST nominal beam parameters (RMS bunch length $\sigma_{z,i}$ is 3 – 4 ps and bunch charge is 0.25 – 3.2 nC). The Particle-In-Cell (PIC) simulation (CST-PS) includes space charge and CSR effect and nonlinear energy distribution over macro-particle data. For Elegant simulations, bunch charge distribution and the beam spectra are mainly investigated with three different bunch charges, 0.25 nC, 1 nC, and 3.2 nC, under two RF-chirp conditions of minimum and maximum energy spreads. The corresponding bunch length for the maximally chirped beam is 2.25, 3.25, and 4.75 ps and the correlated energy spread is 3.1, 4.5, and 6.2 % respectively for bunch charge of 0.25 nC, 1 nC, and 3.2 nC.

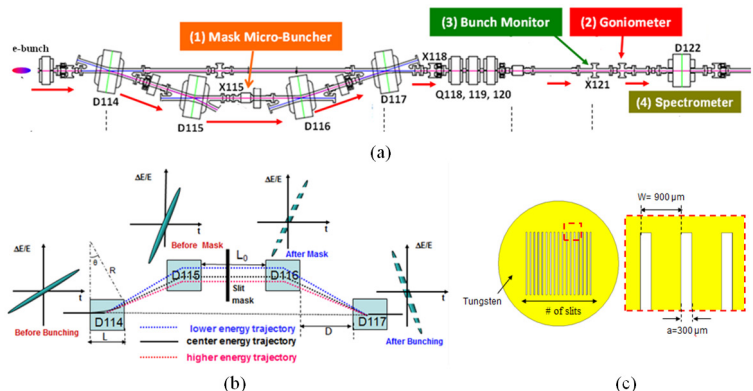

Figure 1: (a) Fermilab-ASTA injector beamline configuration for the slit-mask micro-bunching experiment (b) time-tagged phase-space diagram of bunch modulation process through the masked chicane ($L_0$ = 0.92125 m, $D$ = 0.6825 m, $\alpha$ = 18°, $R$ = 0.95 m) (c) designed slit-mask (CST-PS/Impact-Z model).

## 2. Analytic Design of the Slit Mask

The designed chicane consists of four dipoles and a slit mask with slit spacing, $W$, and aperture width, $a$, is inserted in the middle of the bunch compressor (dispersion region). The configuration of such a masked chicane is shown with the phase-space plots in Fig.1 (b). Before the beam is injected into the masked chicane, a positive linear energy-phase correlation is imposed by accelerating the beam off the crest of the RF wave in the linear accelerator. The chicane disperses and re-aligns the particles with respect to their energies in phase space. The input beam is then compressed and the phase space ellipse is effectively rotated toward the vertical. In the middle of the chicane, the beam is partially blocked by the transmission mask and holes are introduced in the energy-phase ellipse. In the second half of the chicane, the beam is deliberately over-bunched and the beam ellipse is slightly rotated past the vertical. In this step, the linear energy-phase correlation is preserved by over-bunching, accompanied with a steeper phase-space slope. Consequently, the projection of the beam ellipse on the time axis generates density modulations at a period smaller than the grid spacing.

With a masked chicane, one can control the microscopic structure of a bunch under compression by adjusting the grid period and/or by varying the chicane magnetic field. In principle, if a grid period (or slit-spacing) is smaller than a hundred microns, a modulation wavelength of the bunched beam is possibly cut down to a few tens of microns. The beamline for the mask is originally designed with the four dipoles having bending angle of 18°, bending radius $R$ = 0.95 m, and dipole separation $D$ = 0.68 m. The 125 μm thick tungsten mask with a slit-array is designed with period of $W$ = 900 μm and aperture width of $a$ = 300 μm (~ 33 % transparency). The masked chicane and the detailed parameters are illustrated in Fig.1 (b).

According to reference [5] and [16], the longitudinal dispersion, $R_{56}$, and maximum transverse dispersion, $\eta_x$, of a magnetic chicane are given as

$$R_{56} = \gamma \frac{dz}{d\gamma} = -4L\sec\theta + 4R\theta - 2D\sec\theta\tan^2\theta \qquad (1)$$

$$\eta_x = \gamma \frac{dx}{d\gamma} = -2L\tan\theta + 2R(1-\cos\theta) - D\tan\theta(1+\tan^2\theta) \qquad (2)$$

For the FAST 50 MeV chicane, $R_{56}$ = - 0.192 m and $\eta_x$ = - 0.34 m. The bunch-to-bunch spacing (or modulation wavelength), $\Delta z$, is defined by the final bunch length divided by the number of microbunches in a compressed beam [6]. The final bunch length then becomes

$$\sigma_{z,f} = \sqrt{(1+h_1 R_{56})^2 \sigma_{z,i}^2 + \tau^2 R_{56}^2 \sigma_{\delta i}^2} \qquad (3)$$

, where $h_1$ is the first order chirp, $R_{56}$ is the longitudinal dispersion, $\sigma_{z,i}$ is the initial bunch length, $\sigma_{\delta i}$ is the initial un-correlated RMS energy spread, and $\tau$ is the energy ratio. The energy ratio is normally ~ 0.1 at FAST photoinjector beamline. The beam compression is defined by

$$\eta = \sigma_{z,i}/\sigma_{z,f} \qquad (4)$$

After passing through a slit-masked chicane, the number of microbunches of the compressed beam is determined by the transverse beam size at the mask, $\sigma_{x,mask}$, and the slit period, $W$, by

$$N_b = \frac{\sigma_{x,mask}}{W} \qquad (5)$$

The correlated energy spread, $\sigma_\delta$, and transverse emittance, $\varepsilon_x$, normally determine a transverse beam size at the mask by

$$\sigma_{x,mask} = \sqrt{\varepsilon_x \beta_{x,mask} + (\eta_{x,mask}\sigma_\delta)^2} \qquad (6)$$

Where $\beta_{x,mask}$ is the beta function, $\eta_{x,mask}$ is the transverse dispersion at the mask [17, 18], and $\sigma_\delta$ is the correlated energy spread given as

$$\sigma_\delta^2 = \tau^2 \sigma_{\delta i}^2 + h_1^2 \cdot \sigma_{z,i}^2 \qquad (7)$$

The bunch-to-bunch spacing of modulated beam, $\Delta z$, can thus be derived to be

$$\Delta z = W \frac{\sqrt{(1+h_1 R_{56})^2 \sigma_{z,i}^2 + \tau^2 R_{56}^2 \sigma_{\delta i}^2}}{\eta_{x,mask} h_1 \sigma_{z,i}} = W \frac{\sqrt{\left(\sigma_{z,i} + R_{56}\sqrt{\sigma_\delta^2 - \tau^2 \sigma_{\delta i}^2}\right)^2 + \tau^2 R_{56}^2 \sigma_{\delta i}^2}}{\eta_{x,mask}\sqrt{\sigma_\delta^2 - \tau^2 \sigma_{\delta i}^2}} \qquad (8)$$

With the calculated bunch-to-bunch spacing, the bunch length of microbunches can be evaluated by $\sigma_{z,m} = T \cdot \Delta z$, where $T$ (= $a/W$) is the mask transparency (~ 33 % here), assuming the time pattern of the beam is similar to the mask pattern [6].

We examine bunch lengths, compression ratios, transverse beam size at the mask position, and bunch-to-bunch distances with respect to correlated energy spreads, $\sigma_\delta$, for the beam with FAST nominal parameters, as described in Ref. [7], for three different bunch charges, 250 pC, 1 nC, and 3.2 nC. The modulated bunch profiles are calculated with the following condition based on the FAST chicane design parameters: $\sigma_{\delta,i}$ = 0.1 %, $\tau$ = 1, $R_{56}$ = - 0.192 m, and $\eta_x$ = - 0.34 m. As shown in Fig.2 (a), for a beam with small correlated energy spread ($\sigma_\delta$ ~ 0.1 %), the bunch is barely compressed and the final bunch length ($\sigma_{z,f}$) is nearly same as initial bunch length $\sigma_{z,i}$ (= 1.93 mm for 250 pC, 1.95 mm for 1 nC, and 2.56 mm for 3.2 nC). One can see that the compression becomes quickly effective and the bunch length is steeply shortened as $\sigma_\delta$ increases until it reaches 1 %. When $\sigma_\delta$ reaches about 1 – 2 % with $h_1$ = - 1/$R_{56}$, the beam is maximally compressed and the final rms bunch length ($\sigma_{z,f}$) tends to approach $\tau \cdot R_{56} \cdot \sigma_{\delta i}$. Note that further increase of $\sigma_\delta$ renders the beam less compressed and would make the beam rather stretched instead of being compressed. The bunch length via the magnetic chicane is thus mainly governed by an amount of beam energy-spread determined by a beam injection condition with respect to RF-phase. As shown in Fig. 2(b), the compress ratio is apparently in inverse proportion as a final bunch length (rms), which therefore overshot when the beam is maximally compressed. A transverse beam size at the mask position dominantly depends upon transverse dispersion ($\eta_x$) and correlated energy spread ($\sigma_\delta$) since in general they are relatively much larger than the term with beam emittance and beta function, as shown in Fig. 2(c). A number of micro-bunches of a compressed beam is therefore mainly determined by a beam dispersion of the chicane. For the nominal beam parameters with respect to bunch charges [7], the transverse beam size varies from 2 mm up to 18 mm with correlated energy spread 0 – 10 %. With $W$ = 900 μm, the slit-mask will produce $N_b$ = 2 – 20 of microbunches over the beam sizes. Figure 2(c) shows bunch-to-bunch distance (bunch modulation length) with correlated energy spread, $\sigma_\delta$. The analytic calculation points out that the distance becomes ~ 100 μm with correlated energy spread of 1 ~ 2 %. With a 33.3% mask transparency, it is predicted that the ~ 100 μm spaced bunch produces a microbunch length of ~ 33 μm, corresponding to 100 fs in time.

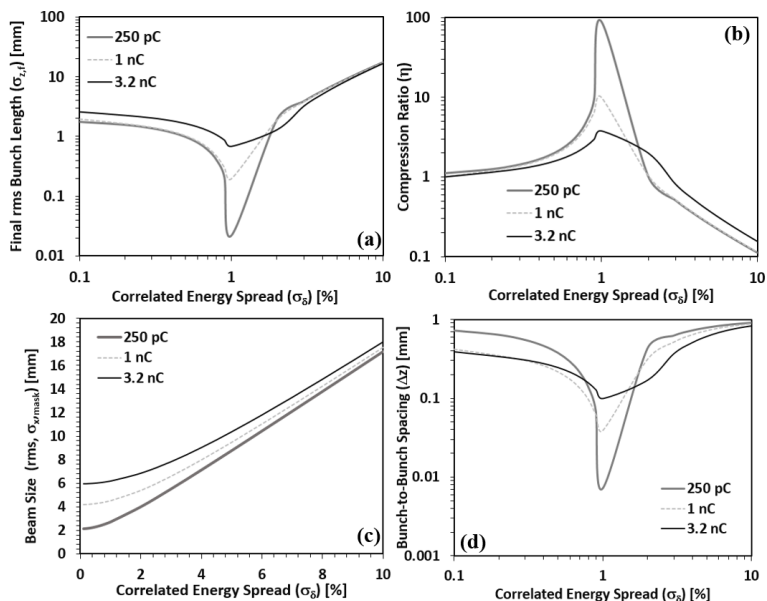

Figure 2: Theoretical assessment of modulated bunch parameters with the ASTA beam profiles [7] (a) final bunch length (rms, $\sigma_{z,f}$), (b) compression ratio ($\eta$), (c) transverse beam size (rms, $\sigma_{x,mask}$) at the slit-mask, and (d) bunch-to-bunch spacing of modulated beam ($\Delta z$) versus correlated energy spread ($\sigma_\delta$) with respect to un-correlated initial energy spreads ($\sigma_{\delta i}$).

In a chicane, bending dipoles are a source of producing CSR along the beamline. However, in general perfectly symmetric dipoles lead the CSRs out of phase (π phase advance) between identical CSR emission points when their optics is properly aligned, which completely cancels successive CSR kicks. Our theoretical analysis is thus based on the linear model free from the CSR effect. However, in practice spatial and other optics constraints break such a symmetry, producing CSR kicks, which induces emittance growth. The CSR-induced emittance growth, proportional to relative rms energy spread per unit length induced by CSR, change a beam size at the slit-mask position, which would affect micro-bunching dynamics of a sliced beam. Therefore, the designed masked chicane is examined with finite-difference-time-domain (FDTD) PIC-simulations to analyze its beam-modulation performance with the CSR effect and also nonlinear beam dynamics on space charge force.

## 3. Simulation Analysis

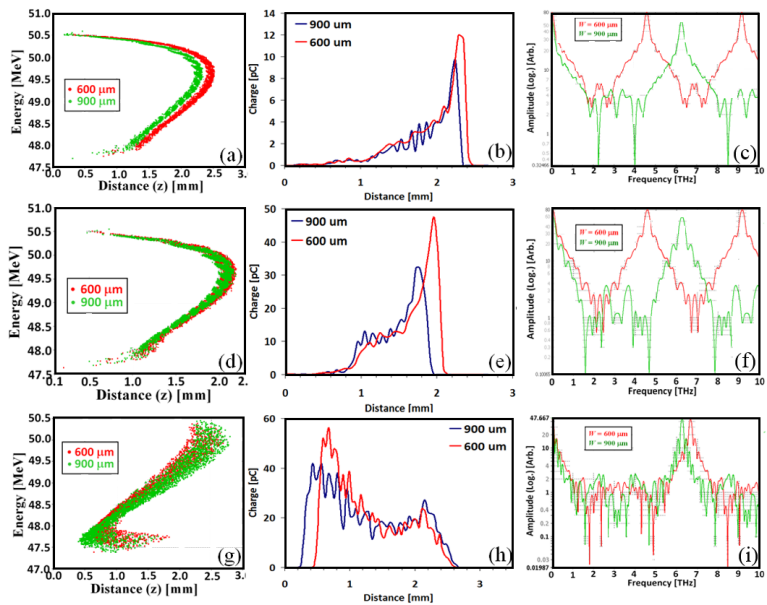

*Figure 3: Simulation results from CST-PS/Impact-Z of the slit-masked chicane with 250 pC (top), 1.0nc (middle) and 3.2 nC (bottom), including energy distribution ((a), (d), and (g)), charge distribution, (b), (e), and (h), in longitudinal distance, and the FFTed spectra, (c), (f), and (i), with $W = 900$ μm (blue) and $W = 600$ μm (red), respectively.*

In order to verify the analytic model, the masked chicane is simulated by CST-PS and Elegant with macro-particle data imported. For CST-PS simulations, 10,000 macro-particles with nonlinear energy distribution are transferred from Impact-Z [21]. For Elegant simulations, macro-particles are imported from a space-charge tracking code, ASTRA [22], which is combined with an extended analysis program called "Shower [14]" to include particle transition effect through a mask material. In order to analyse characteristics of the bunched beam, beam energy distribution, charge distribution, and beam spectrum are monitored at the exit of the chicane. As shown in Fig.3, two different slit-masks with $W = 900$ μm and $a = 300$ μm and with $W = 600$ μm and $a = 200$ μm are modelled with bunch charges of 250 pC, 1.0 nC and 3.2 nC. As theoretically assessed in section-II, the beam is strongly modulated with $W = 900$ μm and ~ 100 μm of modulation length ($\Delta z$), which is consistent with the peak (~ 3 THz) appearing in the beam spectrum. An impact of CSR kicks or wakefields from the bending points on bunch profiles of the compressed beam does not prominently appear in the FDTD simulations. However, the amplitude of beam modulation is noticeably reduced if the slit is replaced with the one with the period of $W = 600$ μm. The slit-mask design with $W = 900$ μm and $a = 300$ μm is thus selected for further investigation with Elegant.

Two different bunching conditions with minimum and maximum energy spreads (on-crest and off-crest with maximum chirp) are examined with Elegant. Also, the simulation analysis includes three different bunch charges (0.25 nC, 1.0 nC and 3.2 nC). For the chirped beam with bunch charge of 0.25 nC, 1.0 nC and 3.2 nC, the rms bunch length is 2.25 ps, 3.25 ps, 4.75 ps and correlated energy spread 3.1%, 4.5% and 6.2% respectively. The phase space of the injected beam with charge 3.2 nC are plotted in Fig. 4, including normalized particle distribution in *x-y* and *x-t* planes and energy distribution (minimum and maximum energy chirp). Apparently, the beam charge profile follows approximately Gaussian distribution (see Fig. 4(c)) and the minimum energy chirp leads to about 1 % spread (see Fig.4 (d)), which is far less than that of the chirped beam with linear energy distribution and maximum correlated energy spread of 6.2% (see Fig.4 (e) ).

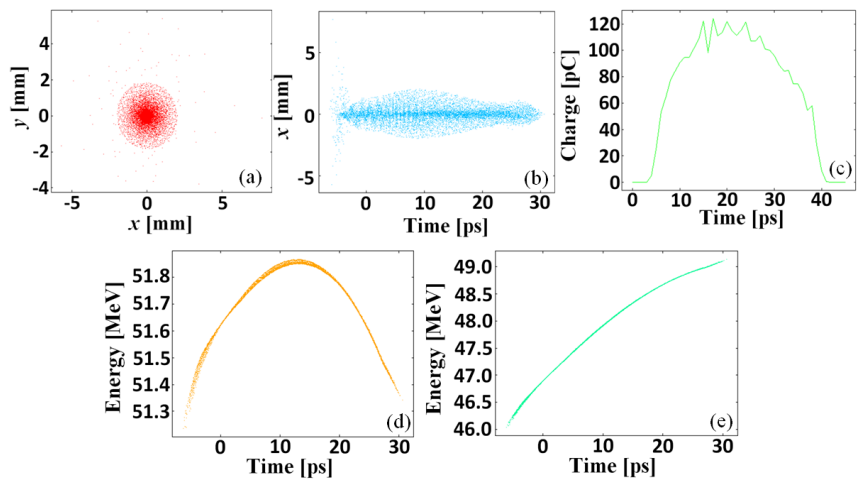

*Figure 4: Phase-space plots (Elegant): (a) x-y, (b) x-t, (c) bunch charge-t, at the entrance of the masked chicane for the beam with 3.2nC bunch charge, Energy distribution with (d) minimum energy spread and (e) maximum chirped beam.*

After passing through the masked chicane, the initial linear energy-time distribution (Fig. 4(e)) is reversed from positive to negative. This conforms to the principle of slit-masked chicane in micro bunch train generation (see ref [5]). The charge distribution for the beam with minimum and maximum energy chirps are shown in Fig. 5 (b) (e) and (i). The beam with minimum energy chirp (red) appears not to carry a modulation pattern in the particle distribution. One can see that under this condition presence of the slit-mask negligibly influences on the beam profile and the chicane behaves as a normal bunch compressor without modulating the beam. On the contrary, the beam modulation under the condition with maximum energy chirp (green) appears much stronger than that with minimum energy spread, as plotted in the normalized charge distribution of Fig. 6. By extracting the abscissa corresponding to peaks of the modulated charge and averaging the distance between adjacent peaks, it is found that modulation wavelengths of 0.25 nC, 1.0 nC and 3.2 nC are about 187 μm, 270 μm and 325 μm, corresponding to bunch lengths of 16 μm, 23 μm and 27 μm, respectively. Note that the linear model presented in section-II predicts 2 - 33 μm of minimum bunch length under the similar condition. Although there are some differences due to approximation in analytic model and some perspectives disregarded in Elegant simulations, those results show the feasibility of ~ 100 fs microbunch generation from the designed chicane. We also notice that the corresponding frequency of the bunch-to-bunch spacing is around 1.6 THz, 1.4 THz, and 1.2 THz, respectively. This is just around the first peak of the frequency spectra of the modulated beam (shown in Fig.5 (c), (f), and (i)). Limited by temporal resolution of each simulation code, the upper limits of the frequency spectra, FFTed from the signals of modulated beams, are set to 10 THz for CST-PS and 15 THz for Elegant, which could be possibly increased with a lower bunch charge.

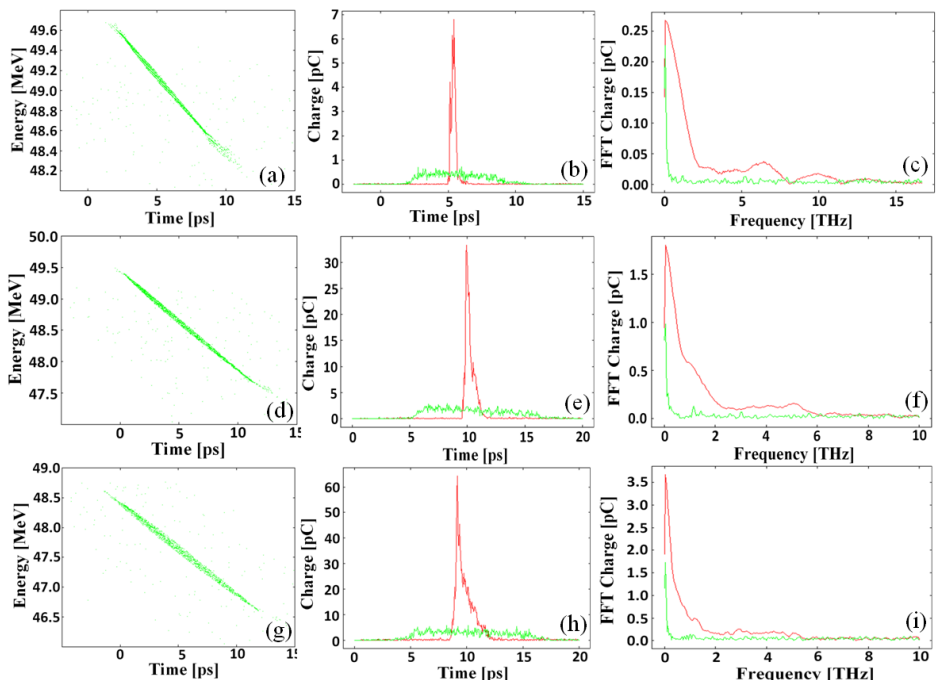

*Figure 5: Simulation results from Elegant/Shower of the slit-masked chicane with 250 pC (top), 1.0nc (middle) and 3.2 nC (bottom), including energy distribution, (a), (d), and (g), charge distribution, (b), (e), and (h), in time, and their FFTed spectra, (c), (f), and (i), with minimum energy spread (red) and maximum energy chirp (green) respectively.*

The difference in density distributions between CST-PS and Elegant (see Fig. 3 and Fig. 5) is attributed to spatiotemporal mesh size and computing algorithm in processing particles colliding with boundaries and slit-mask. In CST-PS and Elegant, bunch charges (y-axis of the graphs) are obtained through multiplying number of macro-particles captured in each equi-spaced mesh along $z$ (CST-PS) or $t$ (Elegant), by the unit charge of the macro-particle. The difference in spatiotemporal mesh size will inevitably lead to a difference in the charge distribution. Furthermore, CST-PS and Elegant have different scheme to deal with particles colliding with boundaries and slit-mask. In CST-PS, the particles colliding with the boundaries and slit mask are absorbed. Together with particle tracking simulations with Elegant, particle transition dynamics in a slit-mask is simulated by program Shower (an interface to Monte Carlo electromagnetic Shower program EGS4 [14]). Despite the similarity of CST-PS and Elegant simulation data, there is still some degree of discrepancy in the bunch charge distribution, which is attributed to several perspectives. In CST-PS, the nonlinear energy distribution induces an asymmetric spatial charge distribution, while charge distribution of Elegant is more symmetric for its linear energy distribution. On the other hand, the injected beams used in CST and Elegant simulation are generated by Impact-Z and ASTRA, based on the linear and nonlinear time-energy correlations, respectively. Also, bunch charges are differently plotted in z-space (CST-PS) and in time (Elegant): the modulated beam charge profile of CST-PS has the opposite pattern to that of Elegant for the beam with charge of 0.25 nC and 1.0 nC. In addition, space charge effect is included in computational algorithm of CST-PS, while it is not considered in Elegant.

Nevertheless, taking into account all the theoretical and numerical analyses, one sees that a properly designed masked chicane can produce a micro-modulated bunch with RMS-bunch length around 100 fs under the optimum beam bunching condition.

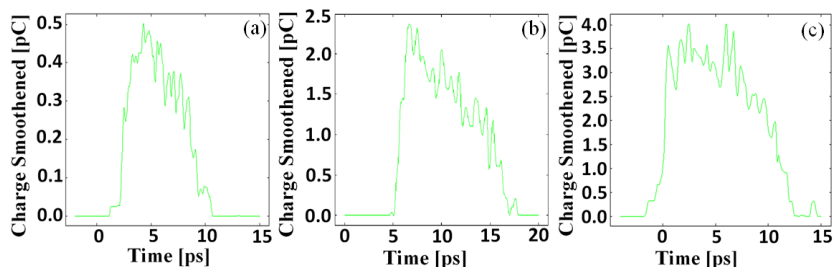

*Figure 6: Normalized charge distribution after the masked chicane for the beam with (a) 0.25 nC, (b) 1.0 nc, and (c) 3.2 nC, which are simulated by Elegant/Shower.*

As described in Section-II, the modulation depth ($\Delta z/\sigma_{z,m}$) is the ratio between the bunch-to-bunch distance ($\Delta z$) and bunch length of a micro-bunch ($\sigma_{z,m}$). The bunch-to-bunch distance ($\Delta z$) is determined by the transverse beam size ($\sigma_{x,mask}$) at a mask position and slit-spacing ($W$), but the micro-bunch length ($\sigma_{z,m}$) is mainly determined by $a$, and $\sigma_{x,mask}$, which is a function of dispersion at the mask position, $R_{56}$, and correlated energy spread, $\sigma_\delta$, if uncorrelated energy spread, $\sigma_{\delta i}$, is small. With a fixed transverse beam size ($\sigma_{x,mask}$) and slit-width ($a$), $\Delta z$ is decreased by decreasing the slit-spacing ($W$), which increases the mask contract ratio, so the modulation depth is decreased. With a very small $W$, the modulation depth would become too small and the beam modulation will disappear. Our simulations indicated that the modulation would not appear if $W$ is smaller than 600 μm with 250 pC bunch charge. The patterns of contrast ratio on beam modulation depth with respect to the slit width and spacing is precited by our theoretical model. A small betatron beam will improve the beam modulation depth. Under the condition in this study, the transverse beam size at the mask is however dominantly determined by correlated energy spread and dispersion of the bunch compressor, the CSR effect would be small on the longitudinal bunch profiles of modulated beam.

## 4. Summary and Conclusion

Since bunch modulation of high brightness beams can significantly improve performance of accelerator-based coherent light sources and high energy linacs, we have investigated a simple way for micro-bunch train generation with a masked chicane, in particular with the bunch compressor at the 50 MeV FAST beamline in Fermilab. The linear model is derived to estimate performance of the designed masked chicane, indicating that the designed slit-mask produces $\sigma_{ms}$ = 33μm long micro-bunches spaced with ~ 100 μm out of $\sigma_t$ = 3 – 4 ps bunch with about 1 – 2 % correlated energy spread.

Numerical analysis with two simulation codes, CST-PS and Elegant, indicates that the beam modulation strongly appears with slit period 900 μm and slit width 300 μm. CST-PS, including nonlinear beam-energy distribution, and space charge effect, results in bunch-to-bunch distance of ~ 100 μm. Also, the simulation shows that the bunch modulation would disappear when the beam is chirped with very small correlated energy spread (on-crest). For the chirped beam, the linear energy-phase space is reversed with strong beam longitudinal charge modulations. The simulation result implies that a mask designed with 900 μm spaced slits (300 μm wide) can split a bunch of 250 pC – 3.2 nC with 3 – 6 % correlated energy spread, while being compressed by a magnetic chicane, into 16 – 27 μm long micro-bunches spaced with 190 – 330 μm modulation length. The simulation data reasonably agree with theoretical analysis of the linear chicane model, which verifies a feasibility of slit-masked chicane to produce a bunch modulation on the order of 100 fs with the beam properly chirped. The result may also offer a useful analytic tool to design/evaluate a short pulse beam shaper and to control six-dimensional beam phase-space for linac-based coherent light sources or multi-bunch high gradient accelerators.

## Acknowledgments

This work was supported by the DOE contract No.DEAC02-07CH11359 to the Fermi Research Alliance LLC. We thank Alex H. Lumpkin, Jayakar C. Thangarj, Darren J. Crawford, Dean R. Edstrom Jr., and Jinhao Ruan of Fermi National Accelerator Laboratory and Philippe R. G. Piot of Northern Illinois University for the helpful discussion and technical supports.